\documentclass[twocolumn]{emulateapj}
\usepackage{epsfig}

\lefthead{Nagy et al. 2011}
\righthead{The Mass-Radius Relation at $z \sim 1.5-3.0$}

\slugcomment{DRAFT: \today}

\begin{document}

\newcommand{\msun}{\ensuremath{\rm M_\odot}}
\newcommand{\msunyr}{\ensuremath{\rm M_{\odot}\;{\rm yr}^{-1}}}\newcommand{\Ha}{\ensuremath{\rm H\alpha}}
\newcommand{\Hb}{\ensuremath{\rm H\beta}}
\newcommand{\lya}{\ensuremath{\rm Ly\alpha}}
\newcommand{\Ntwo}{[\ion{N}{2}]}
\newcommand{\kms}{km~s\ensuremath{^{-1}\,}}
\newcommand{\ztwo}{\ensuremath{z\sim2}}
\newcommand{\zthree}{\ensuremath{z\sim3}}
\newcommand{\feh}{[Fe/H]}

\newcommand{\hst}{{\it HST}-ACS}

\title{The Mass-Radius Relation for Star-Forming Galaxies at $z \sim 1.5-3.0$}
\author{Sarah R. Nagy\altaffilmark{1}, David R. Law\altaffilmark{1,2}, Alice E. Shapley\altaffilmark{1}, Charles C. Steidel\altaffilmark{3}}

\altaffiltext{1}{Department of Physics and Astronomy, University of California, Los Angeles, CA 90095; snagy@ucla.edu, drlaw, aes@astro.ucla.edu}
\altaffiltext{2}{Hubble Fellow.}
\altaffiltext{3}{California Institute of Technology, MS 249-17, Pasadena, CA 91125; ccs@astro.caltech.edu}

\begin{abstract}
We present early results from a {\it Hubble Space Telescope} (HST) WFC3/IR imaging survey of star-forming galaxies in the redshift range $1.5 <  z < 3.0$.
When complete, this survey will consist of 42 orbits of F160W imaging distributed amongst 10 survey fields on the line of sight to bright background QSOs,
covering 65 arcmin$^2$ to a depth of 27.9 AB with a PSF FWHM of 0.18''.
In this contribution, we use a subset of these fields to explore the evolution of the galactic stellar mass-radius relation
for a magnitude-limited sample of 102 spectroscopically-confirmed star forming galaxies ($\langle$SFR$\rangle \sim 30 M_{\odot}$ yr$^{-1}$)
with stellar mass $M_{\ast} \sim 10^{10} M_{\odot}$.  Although the light profile of these galaxies often has an irregular, multi-component morphology, it is typically possible to describe
the brightest component with a Sersic profile of index $n \sim 1$.  The circularized half-light radius $r_{\rm e}$ of the brightest component is on average
$\langle r_{\rm e} \rangle = 1.66 \pm 0.79$ kpc (i.e., $\sim 50-70$\% the size of local late-type galaxies with similar stellar mass), consistent with recent theoretical 
models that incorporate strong feedback from star forming regions.
The mean half-light radius increases with stellar mass and, at fixed stellar mass, evolves
with cosmic time as $\sim (1+z)^{-1.42}$, suggesting that high redshift star forming galaxies may evolve onto the local
stellar mass-radius relation by redshift $z \sim 1$.

\end{abstract}
\keywords{galaxies: high-redshift ---  galaxies: fundamental parameters --- galaxies: structure}

\section{INTRODUCTION}

The redshift range $z \sim 1.5-3.0$ is a critical transition period in the evolution of galaxies.
It is at these redshifts when modern-day galaxies are thought to form the majority of their stars (e.g., Reddy et al. 2008),
fueled by typical star formation rates (SFR) $\sim 30 M_{\odot}$ yr$^{-1}$ (Erb et al. 2006a) comparable to those seen in local starbursts
and driving strong enriched winds into the surrounding intergalactic medium (see, e.g., Steidel et al. 2010).
Previous work has shown that this star formation generally occurs in galaxies with irregular, often multi-component
morphologies (see, e.g., Conselice et al. 2005; Ravindranath et al. 2006; Law et al. 2007; and references therein) at $z\gtrsim2$,
and these galaxies must experience a strong morphological transformation in order to form the Hubble ``tuning fork'' ensemble by $z\sim1.0$ (e.g., Papovich et al. 2005; Ravindranath et al. 2004).

One of the basic properties that is useful for constraining models of galaxy formation and stellar feedback processes is the stellar
half-light radius $r_{\rm e}$, and the evolution
of this quantity with  stellar mass and cosmic time.
Driven by the visible-wavelength surveying efficiency of the ACS camera on board the {\it Hubble Space Telescope} ({\it HST}), 
early efforts to characterize the morphologies of galaxies at $z\sim1.5-3$ (e.g., Abraham et al. 1996; Conselice et al. 2004; Lotz et al. 2006; Papovich et al. 2005; Law et al. 2007)
have generally focused on rest-frame UV emission tracing the regions of active star formation.
While rest-frame UV and rest-frame optical morphology are generally similar for many star forming galaxies
(e.g., Bond et al. 2011), there can be significant morphological differences between the evolved
stellar population (traced by rest-optical emission) and the sites of active star formation (traced by rest-UV emission) for some high-mass galaxies
(e.g., Toft et al. 2005; Law et al. {\it in prep.}) that complicates the establishment of a robust relation between stellar half light radius and stellar mass.

Since rest-frame optical continuum emission is redshifted into the near-IR for galaxies at $z \gtrsim 1$ however,
it has been challenging to constrain $r_{\rm e}$ for a statistically significant sample of galaxies.
Typical $z\sim2$ galaxies are either unresolved or poorly-resolved in seeing limited ground-based imaging, and studies using the {\it HST}/NICMOS camera 
(e.g., Conselice et al. 2011) or
ground-based adaptive optics fed imagers (e.g., Carrasco et al. 2010) have too narrow a field of view to permit efficient surveys of large numbers of galaxies.
Nonetheless, such efforts to characterize the evolution of $r_{\rm e}$ using ground-based instruments and/or
the {\it HST}/NICMOS camera have been made by (e.g.) Papovich et al. (2005), Franx et al. (2008), Toft et al. (2009), van Dokkum et al. (2010),
and Mosleh et al. (2011), generally finding
that galaxies at $z\sim2$ 
were significantly smaller at fixed stellar mass than in the local universe.

With the advent of the new WFC3 camera onboard {\it HST}, it has recently become practical to perform wide-field morphological surveys in the near-IR that
trace rest-frame optical emission from galaxies at $z\gtrsim1$ (e.g., Cameron et al. 2010; Cassata et al. 2010).
Due to a combination of observational limitations and $K$-band selection techniques however, 
these and previous rest-frame optical studies have generally focused upon galaxies with stellar mass $M_{\ast} > 5 \times 10^{10} M_{\odot}$
that are not representative of the bulk of the $z\sim2-3$ star forming galaxy population.  In addition, many studies have relied 
principally upon photometric redshifts, which typically have large uncertainties ($\Delta z/(1+z) \gtrsim 0.06$; van Dokkum et al. 2009) at $z>1.5$, and in individual
cases can sometimes fail catastrophically.

We have therefore undertaken an  {\it HST}/WFC3 imaging survey to map the rest-optical morphology of  a large sample of
optical color selected, {\it spectroscopically confirmed} galaxies in the redshift range $z \sim 1.5-3.0$ with stellar masses in the range
$M_{\ast} \sim 10^{9} - 10^{11} M_{\odot}$.  A complete description of this survey and a full analysis of the morphological properties of these galaxies will
be given by Law et al. ({\it in prep.}): In this contribution we 
present preliminary results  on the mass-radius relation for $z\sim1.5-3.0$ star forming galaxies based on early observations taken as part of our larger imaging survey.
In \S 2 we present the observational data and describe our galaxy sample, outlining our results in \S 3 and discussing their implications for galaxy formation models
in Section 4.
We assume a standard $\Lambda$CDM cosmology in which $H_0 = 71$ km s$^{-1}$ Mpc$^{-1}$, $\Omega_{\rm m} = 0.27$, and $\Omega_\Lambda = 0.73$.

\begin{deluxetable*}{rcccl}
\tablecolumns{4}
\tablewidth{0pc}
\tabletypesize{\scriptsize}
\tablecaption{Mean Circularized Effective Radius $r_{\rm e}$}
\tablehead{
\colhead{} & \colhead{$M_{\ast} = 10^{9.5-10.5} M_{\odot}$} & \colhead{$M_{\ast} = 10^{9.5-10.0} M_{\odot}$} & \colhead{$M_{\ast} = 10^{10.0-10.5} M_{\odot}$} & \colhead{z}}
\startdata
$r_{\rm e}$ & $1.66\pm0.10$ kpc & $1.51\pm0.13$ kpc & $1.84\pm0.14$ kpc & 1.5-3.0\\

$r_{\rm e}$ & $1.90\pm0.18$ kpc  & $1.79\pm0.20$ kpc   & $2.10\pm0.34$ kpc & 1.5-2.0\\

$r_{\rm e}$ & $1.65\pm0.13$ kpc  &  $1.45\pm0.20$ kpc  & $1.80\pm0.17$ kpc & 2.0-2.5\\

$r_{\rm e}$ & $1.22\pm0.12$ kpc  &  $1.01\pm0.14$ kpc  &  $1.50\pm0.13$ kpc & 2.5-3.0\\

$r_{\rm e}/r_{\rm SDSS}$  & $0.61\pm0.04$ & $0.62\pm0.05$ & $0.61\pm0.05$ & 1.5-3.0\\

$r_{\rm e}/r_{\rm SDSS}$ & $0.73\pm0.07$  & $0.74\pm0.09$   & $0.72\pm0.12$ & 1.5-2.0\\
$r_{\rm e}/r_{\rm SDSS}$  & $0.59\pm0.04$  &  $0.59\pm0.08$   &  $0.58\pm0.05$ & 2.0-2.5\\
$r_{\rm e}/r_{\rm SDSS}$  & $0.45\pm0.04$  &  $0.41\pm0.05$  & $0.51\pm0.04$ & 2.5-3.0\\
\enddata
\label{mrtable.tab}
\end{deluxetable*}
	
\section{DATA, SAMPLE, AND ANALYSIS}

The {\it HST} WFC3/IR camera was used to obtain data as part of the Cycle 17 program GO-11694.
When complete, this program will consist of 42 orbits of integration in the F160W
filter divided amongst 14 pointings in 10 different 
survey fields centered on lines of sight to bright background  QSOs ($z \sim 2.7$).
Each of our pointings consists of three orbits containing three
individual exposures of 900 seconds each, for a total of 8100 seconds per pointing.  
In this contribution, we present early results based on the first 6 pointings
(obtained between Oct 2009 and Aug 2010) in the Q1009+29, Q1217+49, Q1549+19, Q1700+64, and Q2343+12 fields, which cover a total area of 28 arcmin$^2$.
Since these fields are distributed widely across the sky we expect cosmic variance in our combined sample to be greatly reduced relative to comparable surveys
over contiguous regions of sky.

These data were reduced using
MultiDrizzle (Koekemoer et al. 2002) with a pixel scale of 0.08 arcsec pixel$^{-1}$ and a pixel droplet fraction of 0.7, 
resulting in the cleanest and narrowest PSF while ensuring that the RMS variation of the final weight map was less than $\sim 7 \%$ across the field of view. The weight map produced by MultiDrizzle was 
used to construct an RMS map, scaling by a correction factor F$_{\rm A}=0.3933$ (see discussion by Casertano et al. 2000) to account for correlation
of the interpixel noise.
We find that the imaging data reach a 
limiting depth of 27.9 AB for a $5\sigma$ detection within a 0.2 arcsec radius aperture, and have a PSF FWHM of $0.18 \pm 0.01$  arcsec
estimated from unsaturated stars.

Our galaxy sample is drawn from rest-UV color-selected catalogs described by Steidel et al. (2003, 2004) and Adelberger et al. (2004).
These catalogs were constructed based on deep ground-based $U_n G {\cal R}$ imaging data. With sizes of $< 1$ arcsec the target galaxies are effectively unresolved
in these ground-based imaging data, and we therefore do not expect our sample to be affected by any significant morphological or surface brightness biases. 
We consider in this work only those 146 galaxies that have been
spectroscopically confirmed (using Keck/LRIS rest-UV spectroscopy) to lie in the redshift range $z = 1.5 - 3.0$, i.e., the
``BM'', ``BX'', and ``LBG'' (strictly ``C'', ``D'', ``M'', and ``MD'') color-selection criteria described by Steidel et al. (2003, 2004).
With an effective wavelength of $\lambda = 15369$ \AA, the F160W filter 
approximately traces from $B$- to $V$-band rest-frame wavelengths over the redshift range of our sample.
All galaxies that have spectroscopic redshifts are detected in the WFC3 imaging data, down to a faint-magnitude limit of  $F160W \sim 25.2$ AB.
Since quantitative morphological measurements become unreliable in these data fainter than $F160W \sim 24.0$ AB however (see discussion by Law et al. {\it in prep.}), we
impose an apparent magnitude requirement that  $F160W \leq 24.0$.
Additionally, we exclude from our sample 6 galaxies which are known to contain AGN or faint QSO on the basis of either photometric or spectroscopic indicators.
This combination of selection criteria results  in a final sample of 102 galaxies (see Figure \ref{histogram.fig}).

$F160W$ magnitudes for the target galaxies were obtained using SExtractor (Bertin \& Arnouts et al. 1996) and corrected for nebular line emission
as detailed by Law et al. ({\it in prep.}).\footnote{In brief, we estimate approximate nebular line fluxes from the UV-derived star formation rate,
and adopt the full amount of the correction as our uncertainty.}
In addition to the deep ground-based $U_n G {\cal R}$ imaging data that forms the backbone of the color-selected galaxy catalog, 
most fields also have ground-based $J$- and/or $K_s$-band imaging, and in some cases Spitzer IRAC and/or MIPS photometry.
These data were combined to obtain stellar mass estimates
by performing  a spectral energy distribution (SED) fitting analysis using Charlot \& Bruzual (2011, {\it in prep.})
stellar population synthesis models with a Chabrier (2003) initial mass function (IMF).
Stellar masses for the star-forming galaxy sample vary from $M_{\ast} = 10^9 - 10^{11} M_{\odot}$, although our sample is 
most well-sampled in the mass range $M_{\ast} = 10^{9.5} - 10^{10.5} M_{\odot}$

Effective (i.e., half-light) radii were obtained using the GALFIT 3.0 program (Peng et al.  2010) to fit a 2D Sersic powerlaw profile
\begin{equation}
\Sigma(r) = \Sigma_e \, {\textrm{exp}} \left[ -\kappa \left( \left (\frac{r}{r_{\rm e}}\right ) ^{1/n} -1 \right ) \right ]
\end{equation}
to the F160W galaxy morphology, using isolated
bright  ($F160W < 20$ AB) but unsaturated stars in each field for the PSF model.  
GALFIT calculates the semi-major axis radius $a$ by default, we convert this to a circularized effective radius
using the formula $r_{\rm e} = a \sqrt{(b/a)}$ (where $b$ is the semi-minor axis).
The PSF FWHM is 0.18 arcsec, or about 1.5 kpc at $z \sim 2$, but since GALFIT deconvolves the PSF it is possible to measure effective radii smaller
than this.  Using the method described by Toft et al. (2007), we find a ${3\sigma}$ resolution limit of 0.073 arcsec, corresponding
to $r_{\rm e, 3\sigma} = 0.62$ kpc at $z = 2$.   Nine galaxies are computed to have effective radii smaller than this value and are thus consistent with a $\sim 9$\% unresolved point source fraction.

Since many galaxies have an irregular, multi-component morphology it is challenging to know how to define the characteristic radius of these systems, and whether it should
describe the size of the brightest component clump or the overall size of the system.
In the present contribution, in galaxies for which
there was more than one component to the light profile (i.e., a single-component Sersic model resulted in a significant residual), 
we used multiple Sersic components to describe the galaxy and adopted the
circularized effective radius of the primary component  (i.e., the brightest in F160W flux)  as the $r_{\rm e}$ of the galaxy.
The primary component is well-defined for most of our galaxies.  In $\sim10{\%}$ of cases, however, the primary and secondary components have comparable magnitudes,
and our stellar masses (which were determined from photometry with angular resolution too poor to distinguish individual clumps) are likely overestimates of the mass of the brightest clump.
As discussed in \S \ref{discussion.sec}, however, our results are largely insensitive to this complication.


\section{RESULTS}
\begin{figure*}[h!]
\begin{center}
\begin{tabular}{c c}
\includegraphics[scale=0.35]{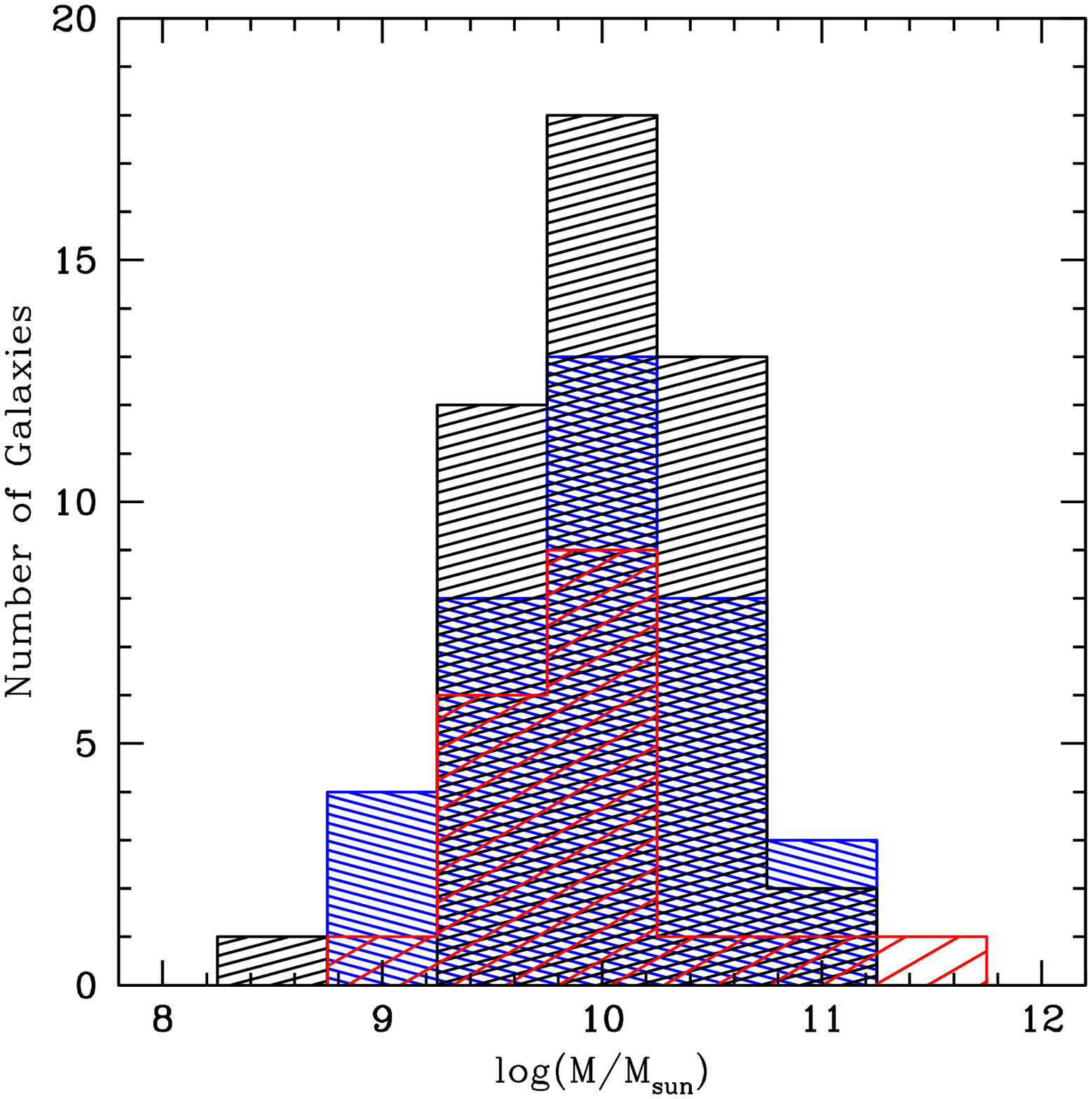} & \includegraphics[scale=0.35]{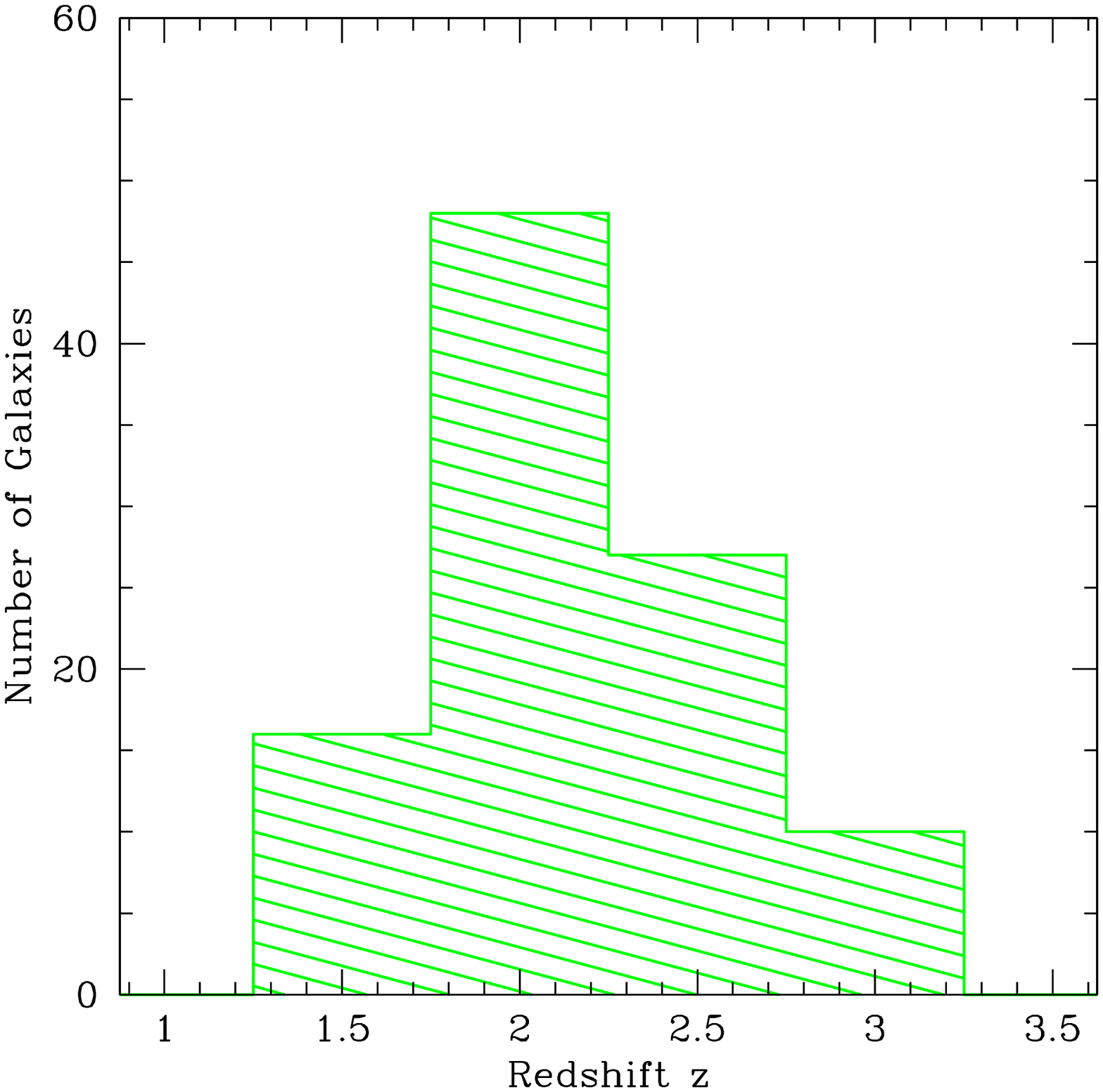}\\
\includegraphics[scale=0.35]{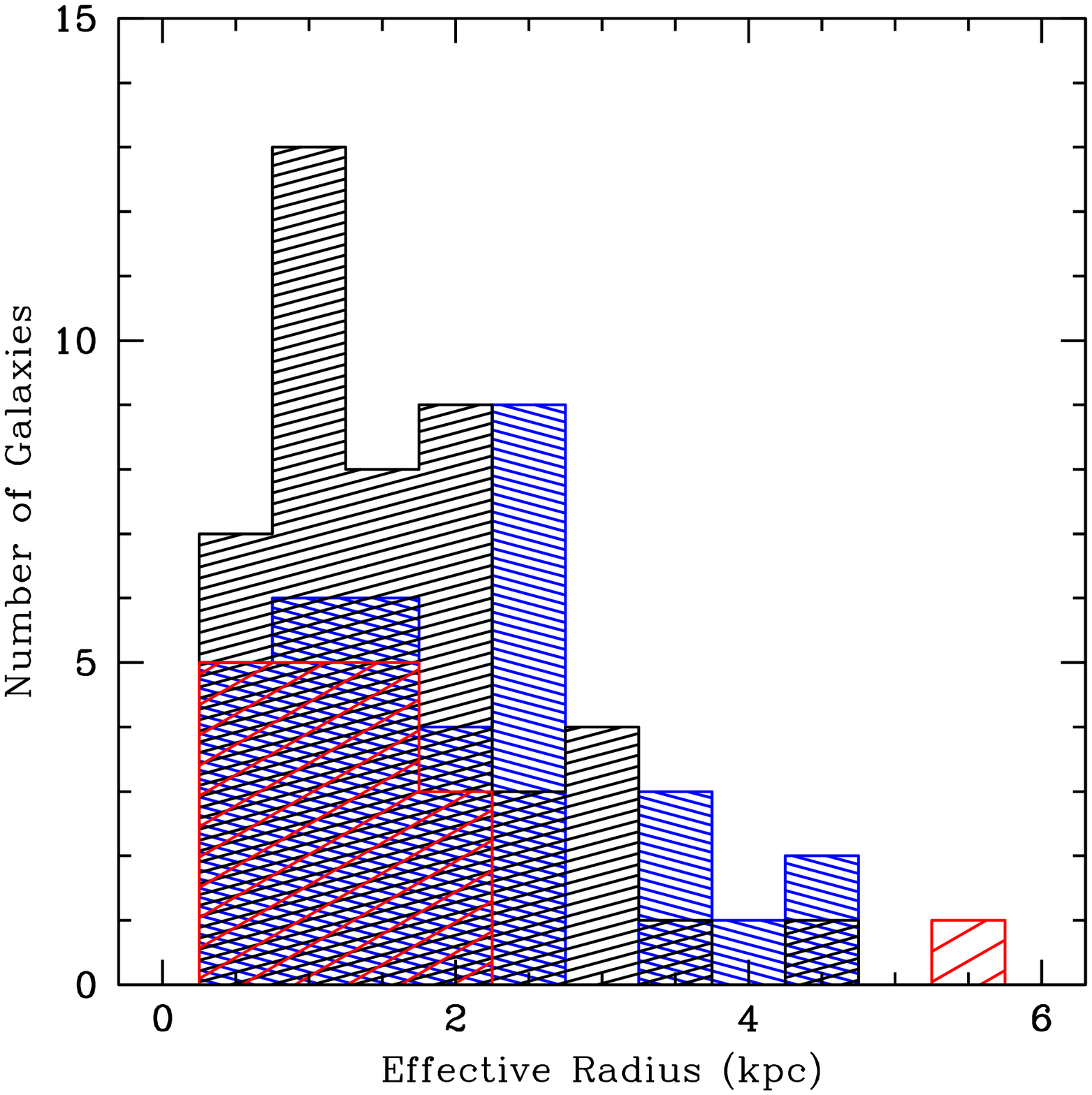} & \includegraphics[scale=0.35]{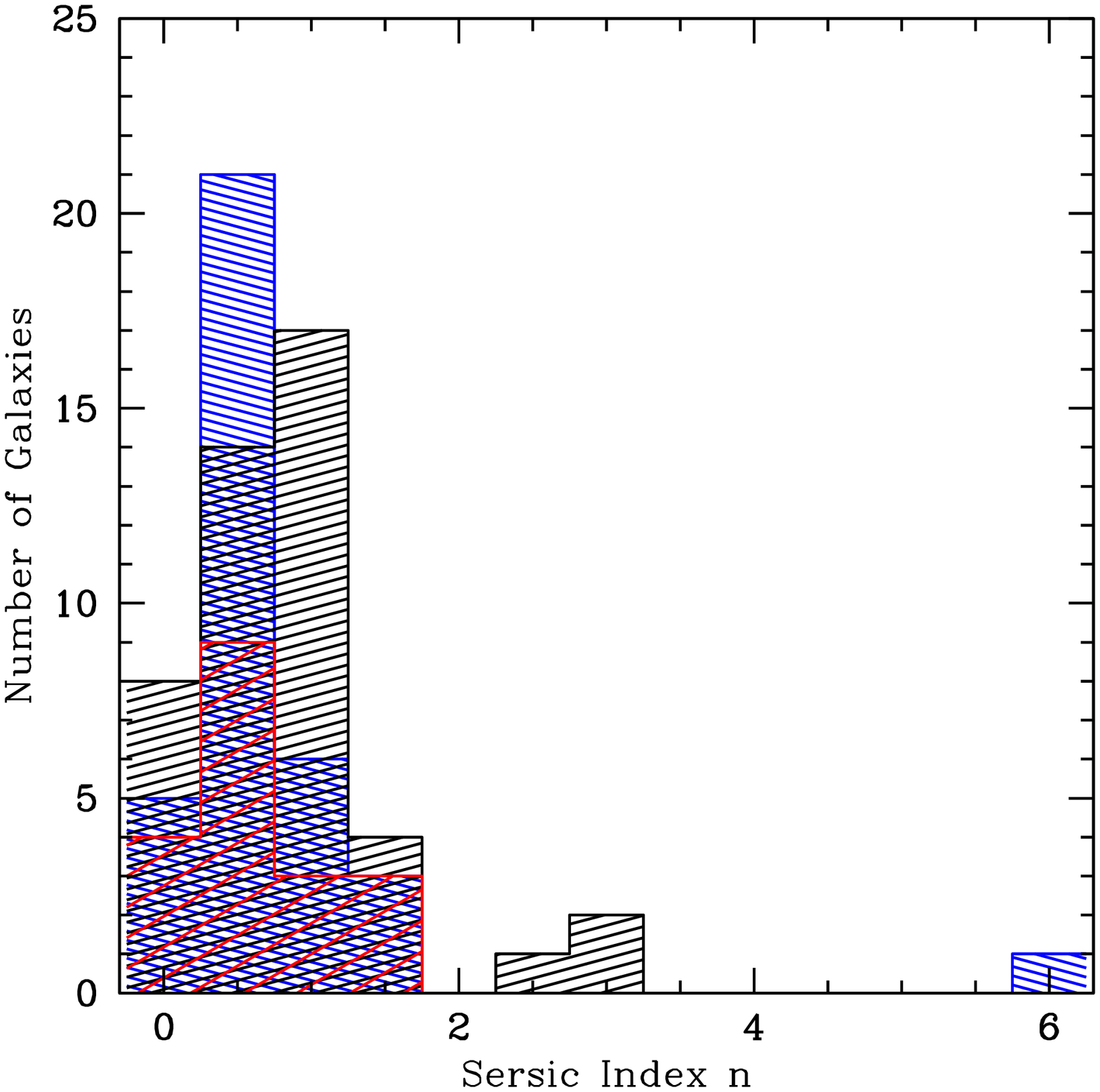}
\end{tabular}
\caption {Histogram of masses log$({M/M_{\odot}})$, redshifts $z$, effective radii $r_{\rm e}$, and Sersic index $n$ for the 102 galaxies in the $z=1.5-3.0$ star forming galaxy population.  The blue/black/red histograms respectively indicate
galaxies in the redshift ranges $z=1.5-2.0$, $z=2.0-2.5$, and $z=2.5-3.0$. The green histogram represents galaxies of all redshifts in the sample.}
\label {histogram.fig}
\end{center}
\end{figure*}

As illustrated in Figure \ref{histogram.fig}, galaxies range in effective radius from unresolved to $r_{\rm e } = 5.5$ kpc, with a Sersic index $n < 2.5$
in all but three cases (two of which had $r_{\rm e}$ values consistent with an unresolved point source).  
Typical stochastic uncertainties in $r_{\rm e}$ and $n$ vary as a function of magnitude and morphological type,
but are generally $\sim$ 3\% and 8\% respectively.
We caution, however, that while the light profile of the primary component of
these galaxies is therefore ``disklike'' in the sense that the light
profile falls off more slowly with radius than for bulge-like profiles ($n = 4$), these are generally {\it not} classical disks in a morphological
or kinematic (e.g., F{\"o}rster-Schreiber et al. 2009; Law et al. 2009) sense.

\begin{figure}
\plotone{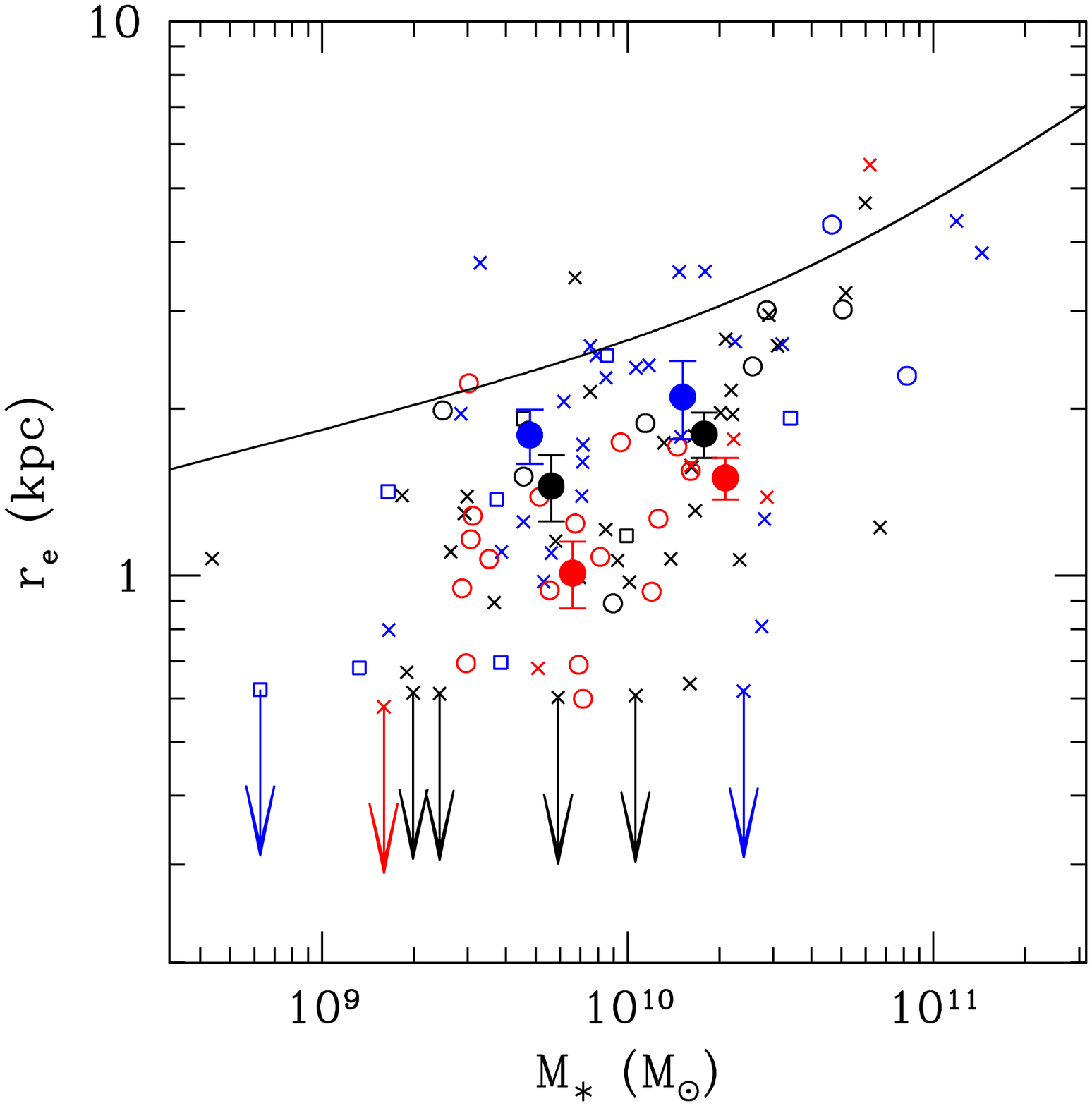}
\caption {Effective circularized radius $r_{\rm e}$ as a function of stellar mass $M_{\ast}$ for the $z=1.5-2.0$ (blue points), $z=2.0-2.5$ (black points), 
and $z=2.5-3.0$ (red points) star forming galaxy samples.
Crosses, open boxes, and open circles represent galaxies selected according to the ``BX'', ``BM'', and ``LBG'' color-selection criteria respectively,
upper limits for unresolved sources are denoted with arrows.
The filled circles and error bars represent the mean value and associated uncertainty in the mean for galaxies in each redshift sample
with stellar masses in the range $M_{\ast} = 10^{9.5-10.0} M_{\odot}$ and $10^{10.0-10.5} M_{\odot}$.
The solid black line indicates the low-redshift relation for late-type galaxies from Shen et al. (2003).}
\label {massrad.fig}
\end{figure}

Plotting the circularized effective radius $r_{\rm e}$ versus stellar mass in Figure \ref{massrad.fig}, we observe that
although there is a large spread in $r_{\rm e}$ for individual galaxies, a mass-radius relation is generally
in place for star-forming galaxies at $M_{\ast} \sim 10^{10} M_{\odot}$ since at least $z \sim 3.0$.  Binning our sample by redshift and stellar mass
(see Table \ref{mrtable.tab}), we calculate that
$\langle r_{\rm e} \rangle = 1.01 \pm 0.14$  ($1.50 \pm 0.13$) kpc for galaxies
in the mass range $M_{\ast} = 10^{9.5-10.0} (10^{10.0-10.5}) M_{\odot}$ respectively at redshift $z = 2.5 - 3.0$, increasing with cosmic time to
$\langle r_{\rm e} \rangle = 1.45 \pm 0.20$  ($1.80 \pm 0.17$) kpc by $z = 2.0 - 2.5$, and to 
$\langle r_{\rm e} \rangle = 1.79 \pm 0.20$ ($2.10 \pm 0.34$) kpc by $z = 1.5 - 2.0$.

\section{DISCUSSION}
\label{discussion.sec}
The radii that we find for galaxies in the redshift $z\sim 2-3$ universe ($r_{\rm e} \sim 1-2$ kpc for galaxies with stellar masses $M_{\ast} \sim 10^{10} M_{\odot}$)
are generally consistent with the theoretical predictions of Sales et al. (2010), and favor their ``WF2Dec'' model in which relatively strong feedback from star forming regions
results in the efficient removal of gas from galaxies via an outflowing wind with velocity $\sim 600$ \kms.  Such peak outflow velocities are generally consistent with observations (see, e.g., Steidel et al. 2010).
Dividing the radii of each of our galaxies by the local value at corresponding stellar mass based on the
late-type (i.e., $n < 2.5$) relation found by Shen et al. (2003) for $\sim 140000$ galaxies in the Sloan Digital Sky Survey (SDSS),
we estimate (see Table \ref{mrtable.tab}) that the radii of $z\sim2-3$ galaxies are on average 
$\langle r_{\rm e}/r_{\rm SDSS} \rangle\sim 50-70$\% the size of comparable-mass galaxies in the low redshift universe.
This is somewhat larger than previous estimates for high-mass ($M_{\ast} > 5 \times 10^{10} M_{\odot}$) star forming and quiescent galaxy populations.
Toft et al. (2009) for instance found that for a sample of 225 galaxies with photometric redshifts from the FIREWORKS catalog, both star-forming
and quiescent galaxies at $z\sim2$ 
were significantly smaller at fixed stellar mass than in the local universe, with $\langle r_{\rm e}/r_{\rm SDSS} \rangle= 0.51 \pm 0.02$
and $\langle r_{\rm e}/r_{\rm SDSS} \rangle= 0.34 \pm 0.02$ respectively.
Similarly,
Franx et al. (2008) found
$\langle r_{\rm e}/r_{\rm SDSS} \rangle= 0.52 \pm 0.06$ for a $K$-selected sample of $z \sim 2$ galaxies  while
van Dokkum et al. (2008) found $\langle r_{\rm e}/r_{\rm SDSS} \rangle= 0.17$ for a $K$-selected sample of galaxies without nebular emission lines.

Toft et al. (2009) postulated that the proximity of $z \sim 2$ galaxies to the local mass-radius relation may be a function of the specific star formation rate (sSFR) of a 
given galaxy, with their Figure 3
demonstrating the increase in $\langle r_{\rm e}/r_{\rm SDSS} \rangle$ from $\sim 0.1$ to $\sim 1.0$ as sSFR increases from $\lesssim 0.2$ Gyr$^{-1}$ (for quiescent galaxies)
to $\sim 1.5$ Gyr$^{-1}$ (for star-forming galaxies).  
Although concerns regarding the evolving mass-to-light ratio as a function of stellar age tend to complicate such trends,
we confirm this statement in the sense that our overall population
has both a large sSFR ($\sim$ 2 Gyr$^{-1}$) and a relatively large mean $\langle r_{\rm e}/r_{\rm SDSS} \rangle = 0.61\pm0.04$,
but note that no such relation is present {\it within} the range of sSFR probed by our star forming galaxy sample.

\begin{figure}
\plotone{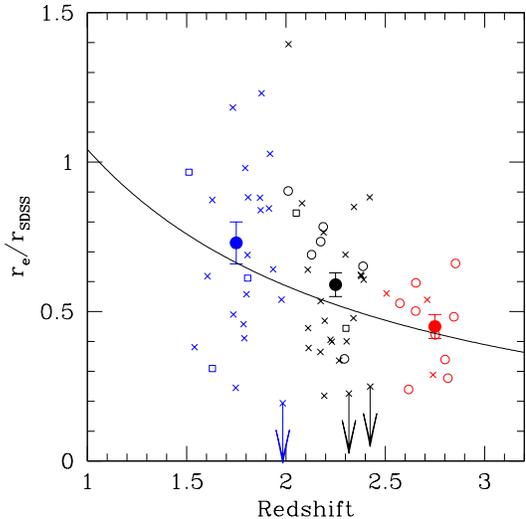}
\caption {Effective circularized radius $r_{\rm e}$ as a fraction of the local relation ($r_{\rm SDSS}$) at a given stellar mass as a function of redshift.  Symbols are as in Figure \ref{massrad.fig};
filled circles represent the mean and associated uncertainty of all galaxies in each of the three redshift ranges.  The solid black line indicates growth of the form $r_{\rm e} \sim (1+z)^{-1.42}$, as derived
from a least-squares fit to the data.}
\label {shenplot.fig}
\end{figure}

The effective radius $r_{\rm e}$ (as a function of stellar mass) of the galaxy population evolves 
significantly closer to the local relation $r_{\rm SDSS}$ with decreasing redshift (see Figure \ref{shenplot.fig}).
Clustering statistics suggest that we are seeing the same population of galaxies at all three redshifts (Adelberger et al. 2005), and
the size of these galaxies increases from 
$\langle r_{\rm e}/r_{\rm SDSS} \rangle = 0.45 \pm 0.04$ in the redshift range $z = 2.5 - 3.0$ to
$0.59 \pm 0.04$ by $z = 2.0 - 2.5$, and $0.73 \pm 0.07$ by $z = 1.5 - 2.0$.
Assuming growth of the form $r_{\rm e} \sim (1+z)^{-\alpha}$ between $z=3.0$ and $z=1.5$, a least-squares analysis gives best-fit $\alpha = 1.42\pm0.50$
(solid line in Figure \ref{shenplot.fig}), consistent with
the $r_{\rm e} \sim (1+z)^{-1.3}$ and $r_{\rm e} \sim (1+z)^{-1.11}$ evolution found by van Dokkum et al. (2010) and Mosleh et al. (2011)
respectively for massive galaxies.
While such an evolution is broadly consistent with classical theories for disk galaxy evolution (e.g., Mo et al. 1998; see also discussion by Papovich et al. 2005, Franx et al. 2008),
it suggests that the $z \sim 2-3$ star-forming galaxy population might be expected to evolve onto the local late-type galaxy relation by $z \sim 1$, in agreement
with Barden et al. (2005), who found little evidence for evolution of the stellar mass-radius relation from $z=1$ to the present day.
We note that our results are robust to our choice to adopt $r_{\rm e}$ of the brightest component as the effective radius for the galaxy while stellar masses are derived from the total
integrated light from all clumps within a given galaxy.  If we instead scale the stellar mass of the galaxy by the fraction of 
F160W flux contained within the primary component (or indeed, if we ignore the comparable-magnitude multi-clump systems entirely)
our derived value of $\langle r_{\rm e}/r_{\rm SDSS} \rangle$ changes by less than 1\%, and the index $\alpha$ of the redshift evolution changes by 
$<0.5\sigma$.

Such a trend might, however, be telling us less about the growth of galaxies within their dark matter halos than about the evolution of the sites of star formation in the young universe.
It is not necessarily surprising that $z \sim 2$ star forming galaxies do not adhere precisely to the local mass-radius relation 
for late-type galaxies since (despite a similar radial index of their rest-frame optical light distribution) they represent drastically different physical systems.
At fixed stellar mass, $z \sim 2-3$ galaxies tend to have much higher gas fractions ($\sim 50$\%, Erb et al. 2006b; Daddi et al. 2010) than late-type disk galaxies in the nearby universe ($\lesssim 20$\%;
Leroy et al. 2008), fueling both high star formation rates $\sim 30 M_{\odot}$ yr$^{-1}$
and star formation surface densities,
 and driving strong gaseous outflows  characteristic of local starburst galaxies.
Similarly, galaxies at $z\sim2-3$ are significantly more morphologically irregular than typical local galaxies (e.g., Abraham et al. 1996; Conselice et al. 2005; Law et al. 2007)
and have much higher gas-phase velocity dispersions $\sigma \sim 70$ \kms (e.g., F{\"o}rster-Schreiber et al. 2009; Law et al. 2009).
As $z$ decreases from $z\sim2$ to $z\sim1$, gas fractions decrease (e.g., Tacconi et al. 2010), morphological properties (e.g., Papovich et al. 2005) 
and kinematics (Wright et al. 2009) begin to resemble the $z \sim 0$ universe,
and it is perhaps unsurprising that we observe evidence for the typical effective radii of galaxies with active star formation beginning to resemble their local values as well.

Indeed, galaxies with properties similar to the $z\sim2$ star forming galaxy sample are not unknown in the low-redshift universe, but are simply much more rare.
The sample of local ($z<0.3$) gas-rich supercompact UV luminous galaxies (ScUVLGs, also sometimes known as LBAs) described by Heckman et al. (2005) for instance are known to
have star formation rates (Hoopes et al. 2007), morphologies (Overzier et al. 2008), and kinematics (Gon{\c c}alves et al. 2010) similar to $z\sim2$ star forming galaxies, and it is perhaps unsurprising
that they appear to have similar rest-optical effective radii as well (Overzier et al. 2010).


Although the morphological sample of galaxies presented here is larger than any previous spectroscopically confirmed $z\sim2-3$
star forming galaxy sample with mass
$\sim M_{\ast} = 10^{10} M_{\odot}$, we caution in closing that it is still relatively small and therefore challenging to explore robustly
the contribution of all possible systematic effects to the measured $r_{\rm e}$.  Such systematics include, but are not limited to, bandshifting throughout the redshift interval,
biases due to multi-component morphologies, biases arising from differences in magnitude between galaxies at different redshifts, etc.
These and other topics will be addressed in detail by Law et al. ({\it in prep.}) using the larger sample of $\sim 400$ galaxies in our full imaging survey.

\section{ACKNOWLEDGEMENTS}

SRN, DRL, and CCS have been supported by grant GO-11694 from the Space Telescope Science Institute.
Support for DRL was also provided by NASA through Hubble Fellowship grant \# HF-51244.01
awarded by the Space Telescope Science Institute, which is operated by the Association of Universities for Research in Astronomy, Inc., for NASA, under contract NAS 5-26555.
AES acknowledges support from the David and Lucile Packard Foundation.

\end{document}